\documentclass[12pt,preprint,natbib]{aastex}
\begin{document}
%\title{Deep Thermal Infrared Imaging of HR 8799 \lowercase{bcde}: New Atmospheric Constraints and Limits on a Fifth Planet}
\title{HR 8799: The Benchmark Directly-Imaged Planetary System}

\author{Thayne Currie\altaffilmark{1}}
%, Marc Kuchner\altaffilmark{1}}
%\altaffiltext{1}{Department of Astronomy and Astrophysics,University of Toronto}
\altaffiltext{1}{National Astronomical Observatory of Japan}
%\altaffiltext{13}{Department of Physics, University of Utah}
\begin{abstract}
HR 8799 harbors arguably the first and best-studied directly-imaged planets.  In this brief article, I describe how the HR 8799 planetary system is a benchmark system for studying the atmospheres, orbital properties, dynamical stability, and formation of young superjovian planets.  Multi-wavelength photometry and spectroscopy show that HR 8799 bcde appear to have thicker clouds than do field brown dwarfs of similar effective temperatures and exhibit evidence for non-equilibrium carbon chemistry, features that are likely connected to the planets' low surface gravities.    Over 17 years of astrometric data constrain the planets' orbits to not be face on but possibly in multiple orbital resonances.   At orbital separations of 15--70 au and with masses of $\approx$ 5--7 $M_{\rm J}$, HR 8799 bcde probe the extremes of jovian planet formation by core accretion: medium-resolution spectroscopy may provide clues about these planets' formation conditions.   Data from the next generation of 30 m-class telescopes should better constrain the planets' orbits, chemistry, gravity, and formation history.
\end{abstract}
%\keywords{planetary systems, stars: early-type, stars: individual: ROXs 42B} 
\section{Introduction}
In March 2008, Christian Marois noticed one (Figure 1, left panel), and then two, faint point sources located at a projected separation of 68 and 38 au from the nearby, dusty A5 star HR 8799.  Follow-up observations in July--September 2008 confirmed that these objects were bound companions and added a third at $\rho$ $\sim$  24 au \citep[][hereafter Ma08]{Marois2008}.   Two years later, \citet{Marois2010} announced the discovery of a fourth companion at $\rho$ $\sim$ 15 au: HR 8799 e (Figure 1, right panel).  
Given the estimated age of the system ($t$ $\sim$ 30 $Myr$, Ma08, Baines et al. 2012), HR 8799 bcde's low luminosities imply masses of  $\approx$ 5--7 $M_{J}$, below the deuterium-burning limit ($\approx$ 13 $M_{J}$) nominally separating planets from brown dwarfs.   Analyses focused on atmospheric modeling \citep[][hereafter Cu11]{Currie2011a}, dynamical stability (\citealt{Marois2010}; Cu11), and formation (\citealt{Kratter2010}; Cu11) likewise corroborate the conclusion that HR 8799 bcde are bona fide planets, not brown dwarfs.

The HR 8799 planetary system resembles a scaled-up version of our outer solar system.  The planets orbit in between a warm dust belt ($r_{inner}$ $\approx$ 6-12 au) and a cold Kuiper belt-like structure at $r_{outer}$ $\approx$ 90--145 au \citep{Su2009,Booth2016}.   Due to HR 8799's higher luminosity, the planets and dust belt populations receive about as much energy as the solar system's gas/ice giant planets and asteroid belt/Kuiper belt receive from the Sun.  

\begin{figure*}[htb]
\begin{center}
\includegraphics[width=0.95\textwidth, trim=0mm 0mm 0mm 0mm,clip]{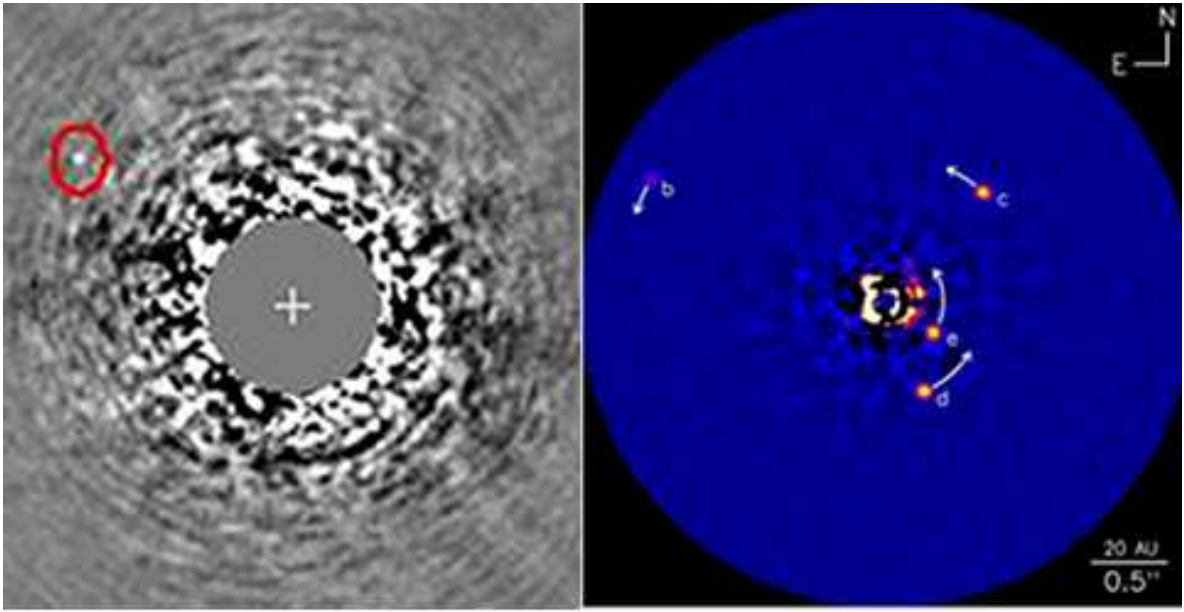}
%\includegraphics[width=0.4\textwidth, trim=0mm 0mm 87mm 0mm,clip]{hr8799_gemkeck.eps}
%\includegraphics[width=0.4\textwidth, trim=87mm 0mm 0mm 0mm,clip]{hr8799_gemkeck.eps}
%\includegraphics[width=0.5\textwidth]{Fig1a.eps}
%\includegraphics[width=0.5\textwidth]{Fig1b.eps}
% OBS: FIGURES MUST BE PS OR EPS FILES
\caption{ (left) The first direct image of an extrasolar planet: detection of HR 8799 b from October 2007 Gemini/NIRI data reduced in March 2008, the first of three planets (HR 8799 bcd) announced by \citet{Marois2008}.  (right) Image of HR 8799 bcde from November 2009 Keck/NIRC2 data \citep{Marois2010} depicting the planets' counterclockwise orbital motion.  The planets' discoveries were enabled by advances in observing and image processing techniques \citep[e.g.][]{Marois2006,Lafreniere2007}.}
%(e.g. Marois et al. 2006; Lafreniere et al. 2007).}
\label{hr8799discovery}
\end{center}
\end{figure*}

HR 8799 harbors arguably not just the \textit{first} directly-imaged planets\footnote{While Fomalhaut b was announced on the same day as HR 8799 bcd and claimed to produce variable, accretion-driven emission at 0.6 $\mu m$ and thermal emission at longer wavelengths \citep{Kalas2008}, later work cast doubt on its existence (Janson et al. 2012) and then showed that instead Fomalhaut b is made visible entirely by circumplanetary dust emission \citep{Currie2012b,Galicher2013}.  Thus, it is on slightly shakier ground and instead, as noted in \citet{Currie2012b}, is likely a ``planet [of unknown mass] \textit{identified by direct imaging} but not a \textit{directly-imaged} planet."   While other planet-mass objects were announced prior to HR 8799 bcd \citep[e.g. 2M 1207 B;][]{Chauvin2004}, their lower mass ratios (compared to the primary) and/or wider separations suggest that they represent the low-mass tail of the substellar mass function.   } but among the best studied ones.  Photometry and/or low-resolution spectroscopy for HR 8799 bcde span 1--5 $\mu m$ (e.g. Cu11; \citealt{Barman2011a, Galicher2011,Zurlo2016}).   HR 8799 bc have 1.4--2.5 $\mu m$ medium-resolution spectroscopy \citep{Konopacky2013,Barman2015}.   After the reported discovery of HR 8799 bcd, multiple studies revealed at least one of the HR 8799 planets from data between 1998 and 2007 \citep[e.g.][]{Lafreniere2009,Metchev2009,Soummer2011,Currie2012a}.  The planets have been imaged by nearly all other 5--8 m telescopes with adaptive optics systems (e.g. Cu11; \citealt{Currie2014a,Ingraham2014,Oppenheimer2013,Zurlo2016}).
% Currie et al. 2014; Ingraham et al. 2014; Oppenheimer et al. 2013; Zurlo et al. 2016).    

This wealth of data makes HR 8799 a benchmark system for studying the atmospheres, orbital properties, dynamical stability, and formation of young superjovian planets.  %providing extensive photometric/spectroscopic measurements needed to clarify HR 8799 bcde's atmospheric properties and 18 years of astrometric coverage that can be used to constrain the planets' orbits and dynamical stability.

% and thus within the planetary mass regime.   
%The four directly-imaged planets around the young A-type star HR 8799 \citep[HR 8799 bcde;][]{Marois2008,Marois2010a} provide a crucial reference point for understanding the 
 %physical properties of gas giants soon after their formation.      
%HR 8799 bcde have best-estimated masses of $\sim$ 5--7 $M_{J}$ \citep{Marois2010a,Currie2011a,Sudol2012} and are located at projected separations of $r_{proj}$ $\approx$ 15--70 $AU$, in between a warm, inner dust belt ($r_{inner}$ $\approx$ 6-12 $AU$) and an outer, Kuiper belt-like structure at $r_{outer}$ $\approx$ 90 $AU$ \citep{Su2009}.   Due to HR 8799's higher luminosity, the planets receive about as much energy as the gas/ice giant planets in our solar system receive from the Sun. Thus, the HR 8799 planetary system may resemble a young, scaled-up version of our own solar system \citep{Marois2010a}.

\section{HR 8799 bcde as a Probe of Young Jovian Planet Atmospheres}
%Photometry and spectroscopy of HR 8799 bcde highlight key, nearly-ubiquitous features of young gas giants several times the mass of Jupiter:  infrared colors and spectral shapes that reveal key differences between (some) young superjovian planets and field brown dwarfs.  Compared to the locus of field brown dwarfs, the planets appear ``red" or ``under luminous" at the shortest infrared wavelengths (i.e. 1--1.6 $\mu m$, Ma08; Cu11).  This trend is due to the planets being cloudier/dustier than field substellar objects at the same effective temperatures (i.e. $T_{\rm{eff}}$ $\sim$ 800--1200 $K$; Cu11).  The clouds may not be uniformly distributed but instead ``patchy"  (Cu11) with 10-50\% of the visible surface covered by thinner clouds (see also Skemer et al. 2012).   At longer wavelengths, some planets show evidence for non-equilibrium carbon chemistry, exhibiting weak to negligible methane absorption at 3.3 $\mu m$  and enhanced $CO$ absorption at 5 $\mu m$ (Hinz et al. 2010; Galicher et al. 2011; Skemer et al. 2012).    Near-infrared spectra for HR 8799 bc reveal molecular species in the planets' atmospheres and additional evidence for non-equilibrium carbon chemistry (Bowler et al. 2010, Oppenheimer et al. 2013, Konopacky et al. 2013; Bonnefoy et al. 2016).    

HR 8799 bcde provide the first glimpse at how the atmospheric properties of young, self-luminous planets compare to both old field brown dwarfs and younger, lower-mass brown dwarfs.   Typical luminosity evolution models (e.g. ``hot start" models) predict that directly-imaged, superjovian (5--10 $M_{J}$) planets between 10 and 100 $Myr$ old cover a temperature range of $\approx$ 600 to 1800 $K$ characteristic of field early L to late T dwarfs \citep[e.g.][]{Baraffe2003,Burrows2006,Stephens2009}.  Coarsely speaking, the L to T transition at $\approx$ 1200--1400 $K$ (for the field) covers a transition from an object with a near-infrared spectrum lacking methane absorption and a cloudy atmosphere to an object with near-infrared methane absorption and weaker/negligible clouds \citep{Saumon2008}.
%(Saumon \& Marley 2008).   
%Previous studies of the youngest and lowest mass brown dwarfs showed hints of departures from the field dwarf sequence, perhaps indicating that the L/T transition is gravity (and thus likely mass/age) dependent (e.g. Luhman et al. 2007).  
 For ages of $\sim$ 30 $Myr$ and masses of 5--7 $M_{J}$, standard luminosity evolution models predict that the HR 8799 planets should, if like field objects, have temperatures of $T_{eff}$ $\approx$ 850--1100 $K$ characteristic of mid/late field T dwarfs \citep{Stephens2009}.
 % (Stephens et al. 2009).  

Ma08 found hints of differences between the HR 8799 planets' infrared (IR) colors and the field L/T dwarf sequence.   Subsequent studies that focused on a wider range of HR 8799 planet colors reveal clear departures from the field sequence (\citealt{Bowler2010}; Cu11).
Compared to mid/late T dwarfs, HR 8799 b(cd) are up to 2.5 (1.5) magnitudes redder (Cu11).  Generally speaking, HR 8799 bcde appear to lie on a reddened extension of the L dwarf sequence past the field L/T transition to fainter magnitudes, a sparsely populated region whose other members are almost unanimously young, low (planetary) mass, and low gravity \citep[e.g.][]{Bonnefoy2016}.
% Bonnefoy et al. 2016).   

The HR 8799 planets' IR spectral shapes likewise reveal differences with field objects but (imperfect) similarities with young, dusty, and low-gravity/mass brown dwarfs.     Although formally no object appears well matched to HR 8799 bc's combined near-IR spectra and thermal IR photometry,  reddened versions of the youngest T0 dwarf spectra reproduce their spectra well \citep{Bonnefoy2016}.   HR 8799 de appear best matched by particularly red and dusty/low gravity L6-L8 dwarfs which likewise deviate from the field sequence.  However, fitting near IR and thermal-IR data simultaneously remains challenging.

As shown in Cu11, the HR 8799 planets appear different than field brown dwarfs of the same effective temperatures in large part because they have thicker clouds (Figure 2).   
%For a given $T_{eff}$, thick clouds translate into a hotter temperature-pressure profile (e.g. Madhusudhan et al. 2011; Marley et al. 2012).   
Thicker clouds change the optical depth profile as a function of wavelength, making it more uniform in and out of major molecular opacity sources (e.g. water) since the $\tau$ = 1 surface is achieved at a more uniform altitude.   The planet spectrum appears redder and more blackbody like.    Furthermore, the clouds may be non-uniformly distributed or  ``patchy"  (Cu11) with 10-50\% of the visible surface covered by thinner clouds/cloudless regions.
\begin{figure}[htb]
\begin{center}
\includegraphics[width=0.85\textwidth, trim=10mm 3mm 5mm 5mm,clip]{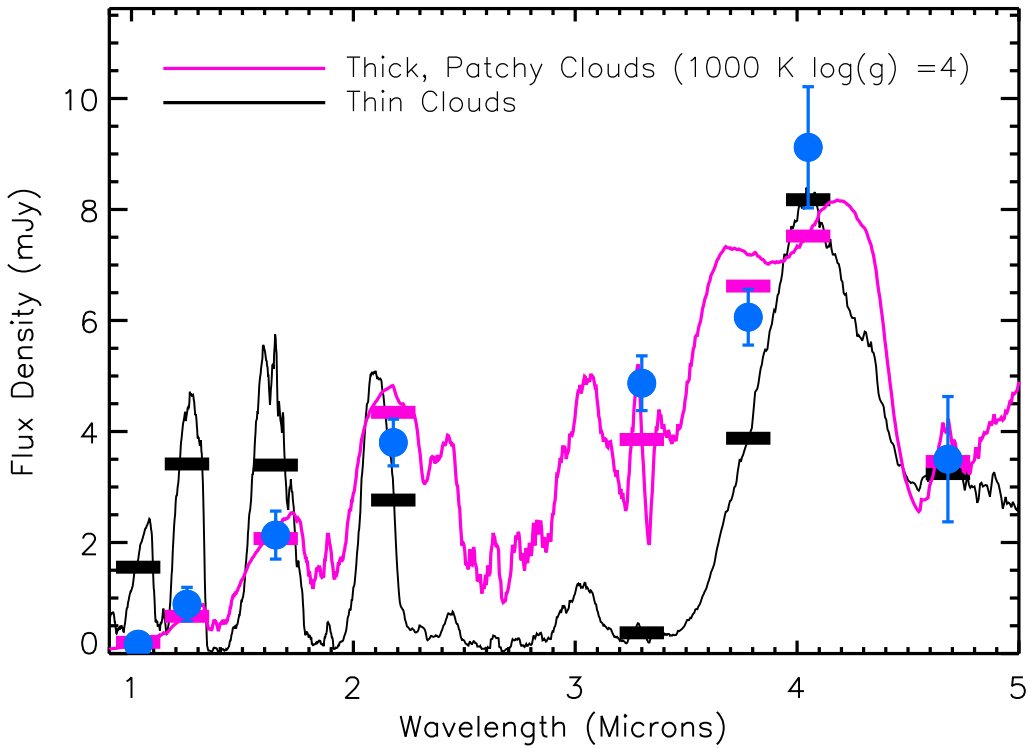}
%\includegraphics[width=0.4\textwidth, trim=87mm 0mm 0mm 0mm,clip]{hr8799_gemkeck.eps}
%\includegraphics[width=0.5\textwidth]{Fig1a.eps}
%\includegraphics[width=0.5\textwidth]{Fig1b.eps}
% OBS: FIGURES MUST BE PS OR EPS FILES
\caption{ Modeling HR 8799 d.  Thin cloud models (black) more appropriate for field T dwarfs overpredict (underpredict) HR 8799 bcde's brightness in the near (mid) IR.  Those invoking thick, patchy clouds (90\% very thick, 10\% moderately thick) provide a far better match (Cu11).  }
\label{hr8799atmos}
\end{center}
\end{figure}
Additionally, at least some HR 8799 planets show clear evidence for non-equilibrium carbon chemistry \citep{Barman2011a,Galicher2011,Skemer2012}.  In addition to the absence of strong $CH_{4}$ absorption in HR 8799 b's low-resolution near-IR spectrum, at longer wavelengths at least some of the planets exhibit weak to negligible methane absorption at 3.3 $\mu m$  and enhanced $CO$ absorption at 5 $\mu m$ \citep{Galicher2011,Skemer2012}.    Medium-resolution near-IR spectra for HR 8799 bc reveal molecular species in the planets' atmospheres and additional evidence for non-equilibrium carbon chemistry \citep{Konopacky2013, Barman2015}.

The planets' low surface gravities (and, thus, their youth and low mass) explain both thick clouds and non-equilibrium carbon chemistry.  Lower gravities (log(g) $\sim$ 4 instead of $\sim$ 5 for field objects; Cu11, Konopacky et al. 2013) yield temperature-pressure profiles more characteristic of hotter, cloudier L dwarfs \citep{Madhusudhan2011}.   Lower gravities also move the depth at which carbon-based chemical reactions (i.e. CO+3H$_{\rm 2}$ $<$--$>$ CH$_{\rm 4}$+ H$_{\rm 2}$0) are quenched deeper in the atmosphere, resulting in an overabundance of CO \citep{Barman2011a,Marley2012}.
%(Barman et al. 2011a; Marley et al. 2012).  %Typical values for eddy diffusion coefficients describing mixing are comparable to those deep in the atmosphere of Jupiter (Barman et al. 2015).
% for 106-108 cm2 s-1
Other young, directly-imaged 5--13 $M_{\rm J}$ planetary-mass objects with a range of temperatures also show evidence for thicker clouds and/or non-equilibrium carbon chemistry \citep[e.g. ROXs 42Bb and 2M 1207 B;][]{Currie2014b, Barman2011b}, although the lowest mass, coldest and oldest imaged planets have very different spectra \citep{Kuzuhara2013, Macintosh2015}.

\section{The Orbits of HR 8799 bcde}
HR 8799 bcde's orbits provide crucial input for studying the configuration of multi-planet systems and a limit on the planets' masses.  Soon after HR 8799 bcd were announced, \citep{Fabrycky2010} noted that for nominal face-on orbits and nominal masses, the system would be dynamically unstable in 0.1 $Myr$, far less than the system age.   Placing these planets in a 4:2:1 orbital resonance makes the system dynamically stable for tens of $Myr$ up to masses of 10--20 $M_{J}$.  However, a 4th planet at 15 au makes dynamical stability more challenging, favoring masses of 5 (7) $M_{J}$ or less for HR 8799 b (cde) and precluding masses above 13 $M_{J}$ (Marois et al. 2010; Cu11).   Assuming different orbital properties (e.g. inclined orbits) further lowers the maximum allowable planet masses \citep[e.g. less than 10 $M_{J}$][]{Sudol2012}.  
\begin{figure}[htb]
\begin{center}
\includegraphics[width=0.85\textwidth, trim=0mm 1mm 0mm 2.5mm,clip]{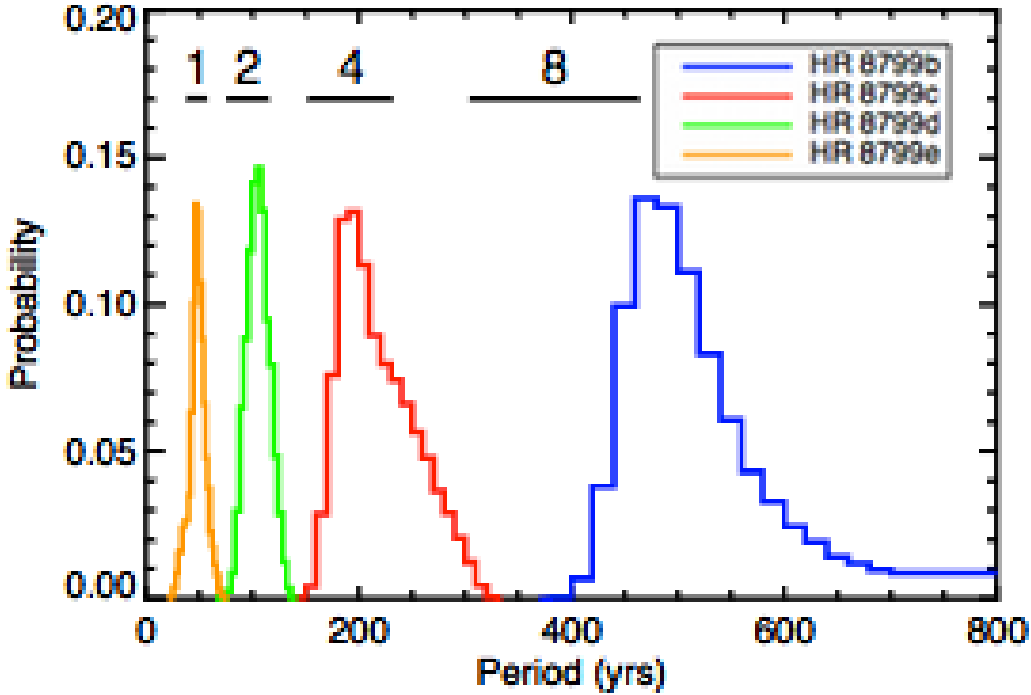}
%\includegraphics[width=0.4\textwidth, trim=87mm 0mm 0mm 0mm,clip]{hr8799_gemkeck.eps}
%\includegraphics[width=0.5\textwidth]{Fig1a.eps}
%\includegraphics[width=0.5\textwidth]{Fig1b.eps}
% OBS: FIGURES MUST BE PS OR EPS FILES
\caption{Range of orbital periods for different orbital solutions from \citep{Konopacky2016}, showing that HR 8799 bcde may be in an 8:4:2:1 resonance.}
\label{hr8799resonance}
\end{center}
\end{figure}

HR 8799 bcde are not in face-on orbits but instead are inclined between 20 and 45$^{o}$ from face-on \citep{Soummer2011,Currie2012a,Pueyo2015,Konopacky2016}, similar to HR 8799 A's rotation axis and the inclination of the star's cold debris disk \citep{Reidemeister2009, Booth2016}.    Circular orbits are formally consistent with the data for all planets \citep{Currie2012a, Konopacky2016}.   HR 8799 bcd's allowable orbits include a stabilizing 4:2:1 resonance, a possible outcome of the planets' formation/migration histories (Figure 3, Konopacky et al. 2016; Gozdziewski \& Migazewski 2014).  However,  it is unclear whether a 2:1 or 3:2 resonance is favored between HR 8799 d and e \citep{Zurlo2016}.

The planets' relative inclinations are uncertain.  \citet{Currie2012a} argue that coplanar orbits comprise a small fraction of acceptable orbits drawn from 1998-2010 measurements but are still possible; \citet{Pueyo2015} suggest that the planets' orbits are not coplanar.   While Konopacky et al. (2016) find no evidence for non-coplanarity from a set of self-consistently reduced data sets originating from the same telescope/instrument configuration, they do not consider astrometry for HR 8799 d (bc) prior to 2007 (2004), measurements largely responsible for favoring non-coplanarity.   
%At least HR 8799b and c must be in
%an orbital resonance for the system to be dynamically stable (Fabrycky \& Murray-Clay 2010).  
%\citep{Galicher2014, Currie2013,Currie2014a,Barman2011,Skemer2011}. 

\section{Formation}
HR 8799 bcde are important tests of jovian planet formation.   Forming 5--7 $M_{J}$ planets at 25--70 au in situ by core accretion is extremely difficult, consistent with the fact that HR 8799-like planetary systems are extremely rare (e.g. Galicher et al. 2016).   The mass ratios and separations of the HR 8799 planets are contiguous with the population of radial-velocity/transit detected planets, at least some of which formed by core accretion (Cu11;\citealt{Kratter2010}).   Nevertheless, uncommon conditions -- a particularly massive disk, very rapid and efficient build up of protoplanetary cores, and/or scattering of massive cores to wider separations prior to runaway gas accretion (e.g. \citealt{KenyonBromley2009}; Cu11; \citealt{Lambrechts2012})-- are likely required to explain HR 8799 bcde.

Atmospheric chemistry may also provide clues to HR 8799 bcde's formation.   If the planets formed by disk instability, their atmospheres should have solar C/O ratios; if they formed by core accretion near their current locations with a modest intake of solids, they should have enhanced C/O \citep{Oberg2011}.   While HR 8799 c may have an enhanced C/O ratio, HR 8799 b's C/O ratio is less constrained \citep{Konopacky2013, Barman2015}.  
\section{Future Prospects}
Many challenges remain in better understanding the atmospheres, orbits, and formation of the HR 8799 planets.
For example, HR 8799 b remains a sort of ``Kobayashi Maru" test for planet atmosphere models\footnote{Portrayed in Star Trek II: The Wrath of Khan (and in the 2009 franchise reboot), the \textit{Kobayashi Maru} test is simulation in which Starfleet cadets captain a starship to rescue a freighter and are subsequently ambushed by three Klingon battlecruisers.  It is designed such that saving the ship is impossible, thus testing how cadets deal with a "no-win scenario".  James T. Kirk nevertheless beat it by ``changing the rules of the game" (cheating) to allow the ship to be saved.  We cannot similarly change the atmospheres of extrasolar planets so that our models fit them.}, as not a single one has reproduced all of the planet's spectrophotometry while yielding parameters (e.g. mass, radius) that are physically plausible and consistent with other constraints (planet cooling models; dynamical stability limits).  Likely (partial) solutions to this problem include updated near-IR opacities and a better understanding of clouds at low gravities.  By 2019, we will have astrometric points for roughly 20\% of HR 8799 d and e's orbit.  New data combined with a self-consistent reanalysis of older astrometry will be needed to better constrain key orbital properties (e.g. resonances, eccentricities, and coplanarity).
%which orbital resonances they reside in and whether or not the planets are on circular, coplanar orbits.  

The next generation of extremely large telescopes will provide powerful probes of the HR 8799 planets' chemical abundances, gravity, and formation history.  For example, the \textit{IRIS} integral field spectrograph on the \textit{Thirty Meter Telescope} covers 0.8--2.5 $\mu m$ and should be capable of providing high signal-to-noise spectra of all four planets at $R$ = 4000--8000, building upon previous Keck/OSIRIS studies of HR 8799 bc at $H$ and $K_{\rm s}$ bands \citep{Konopacky2013, Barman2015}.  Such data should resolve multiple gravity sensitive lines and better determine abundances of multiple species.  As a result, we may better constrain the C/O ratio and formation environment for planets of comparable mass from 15 au to 70 au.

{}
\end{document}